\documentclass[journal]{IEEEtran}

\usepackage{bbm}
\IEEEoverridecommandlockouts

\makeatletter
\def\ps@headings{%
\def\@oddhead{\mbox{}\scriptsize\rightmark \hfil \thepage}%
\def\@evenhead{\scriptsize\thepage \hfil \leftmark\mbox{}}%
\def\@oddfoot{}%
\def\@evenfoot{}}
\makeatother 

\usepackage{times}
\usepackage{color}
\usepackage{algorithm}
\usepackage{algorithmic}
\usepackage{subfigure}
\usepackage{graphicx}
\usepackage{latexsym}
\usepackage{amssymb}
\usepackage{amstext}
\usepackage{amsmath}
\usepackage{textcomp}
\usepackage{bm}
\usepackage{url}

\def\BibTeX{{\rm B\kern-.05em{\sc i\kern-.025em b}\kern-.08em
    T\kern-.1667em\lower.7ex\hbox{E}\kern-.125emX}}
\newcommand{\kETAL}     {{\em et~al. }}      

\title{\LARGE Wireless Communication  and Networking Technologies for Smart Grid: Paradigms and Challenges } \normalsize

\author{Xi Fang,  \emph{Student Member, IEEE}, Dejun Yang, \emph{Student Member, IEEE},  and Guoliang Xue, \emph{Fellow, IEEE}\\Arizona State University, Tempe, AZ, USA
}

\date{}
\begin{document}

\maketitle \thispagestyle{empty}

\begin{abstract}

Smart grid, regarded as the next generation power grid, uses two-way
flows of electricity and information to create a widely distributed
automated energy delivery network.
In this work we present our  vision on smart grid from the
perspective of wireless communications and networking technologies.
We present   wireless communication and networking paradigms for
four typical scenarios in the future smart grid and also point out
the research challenges of the wireless communication and networking
technologies used in   smart grid.

\end{abstract}

\section{Introduction}
The term \emph{grid} is traditionally used for an electricity
delivery system that may support all or some of the following four
operations: electricity generation, electricity transmission,
electricity distribution, and electricity control \cite{Fang:Smart}.

By using two-way flows of electricity and information, smart grid,
an enhancement of the traditional power grid,   attempts to   create
an automated, distributed, and advanced energy delivery network.
This enhanced grid is expected to provide distributed power
generation, self-monitoring, self-healing, adaptive and islanding
microgrid, pervasive control, and various customer choices.

In order to realize these functions,  an advanced information and
communication system underlying the smart grid will play a critical
role.
For example, in the peak period, the electric utility notifies the
users the real-time price so as to convince them to reduce their
power demands.
Therefore, the total demand profile full of peaks can be shaped to a
nicely smoothed demand profile.
This can help electric utility reduce overall plant and capital cost
requirements.
In order to realize this, an information communication network,
which can guarantee the realtime price notification, is required.
Another example is that to realize grid self-healing requires   a
widely deployed monitoring system.
The grid status information obtained by this monitoring system
should be sent to the controller in a timely manner.
Suppose that  a medium voltage transformer failure event occurs in
the smart grid.
This failure can be detected  by the monitoring system and then
reported to the controller promptly.
Therefore, the smart grid  automatically changes the power flow and
recovers the power delivery service.

%
In this article, we focus on the wireless communication and
networking technologies, which may be used in the future smart grid.
Although Xi \kETAL \cite{Fang:Smart} did a comprehensive survey on
the smart grid, in this work we will refine the exploration of smart
grid and present  our  vision for smart grid from the perspective of
wireless communication  and networking technologies.
We present  four wireless communication and networking paradigms for
typical scenarios in the future smart grid and also point out the
research challenges of the communication and networking technologies
used in the smart grid.

The rest of this article is organized as follows. In Section
\ref{sec:overview},  we overview the basic concept of the smart
grid.
Then, we  vision  the wireless communication and networking
paradigms for four important scenarios  in Section
\ref{sec:paragdims}    and describe some   research challenges in
Section \ref{sec:challenge}.
Finally, we  conclude this article in Section \ref{sec:conc}.

\section{Overview of Smart Grid} \label{sec:overview}

\begin{figure}[h]
\centering
\includegraphics[width=0.5\textwidth]{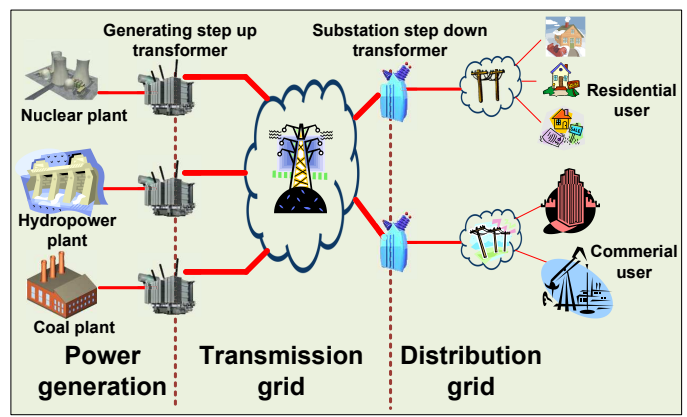}
\vspace{-0.1in} \caption{ 
An Example of the Traditional Power Grid} 
\label{fig:pg_struct}
\end{figure}

A traditional power grid is unidirectional in nature.
Fig.\ref{fig:pg_struct} shows an example of the traditional power
grid.
Electricity is usually generated at  central power plants by
electromechanical generators, primarily driven by the force of
flowing water or heat engines fueled by chemical combustion or
nuclear fission.
These power generating plants are often quite large and located away
from heavily populated areas.
The electric power generated by these plants is  stepped up to a
higher voltage for transmission on a \emph{transmission grid}.
The transmission grid moves the power over long distances to
substations.
Upon arrival at the substation, the power is stepped down  to a
distribution level voltage.
As the power exits the substation, it enters the \emph{distribution
grid}.
Finally, upon arrival at the service location, the power is stepped
down again from the distribution voltage to the required service
voltage(s).

\begin{figure}[h]
\centering
\includegraphics[width=0.5\textwidth]{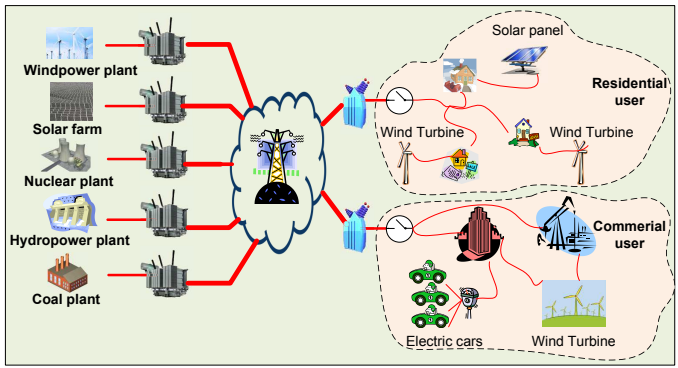}
\vspace{-0.1in} \caption{ 
An Example of the Smart Grid} 
\label{fig:sg_struct}
\end{figure}

In contrast, the smart grid uses two-way flows of electricity and
information to create an automated and   distributed advanced energy
delivery network.
In a smart grid, the energy generation and   delivery is more
flexible.
Fig.\ref{fig:sg_struct} shows an example of a smart grid.
The distribution grid may also be capable of generating electricity
by using distributed   energy generations, such as wind turbines and
solar panels.
Note that the transmission grid could also have large-scale
distributed generators (e.g. wind farm) connected.
Since now    power can be generated by users, we can group these
distributed energy generators and loads as a \emph{microgrid}
\cite{Fang:Smart}.
The microgrid normally operates connected to a traditional power
grid (a \emph{macrogrid}).
The single point of common coupling with the macrogrid can be
disconnected.
The microgrid can then function autonomously  without obtaining
power from the electric utility in the macrogrid.
Thus, the multiple distributed generators and the ability to isolate
the microgrid from a larger network in disturbance will provide
highly reliable electric power supply.
The smart grid also takes advantage of the plug-in hybrid electric
vehicles (PHEVs). PHEVs can communicate with the grid  and deliver
electricity into the grid at peak electricity usage times, if they
are parked and connected to the grid.
These vehicles can then be recharged during off-peak hours at
cheaper   rates.

Due to the complicated power delivery pattern is being used, the
information delivery system should also be updated.
Broadly stated, by utilizing modern information technologies, the
smart grid could respond to events that occur anywhere in the grid,
such as power generation, transmission, distribution, and
consumption, and then adopt the corresponding strategies or
behaviors.

\section{Wireless Communication and Networking Paradigms in Smart Grid}
\label{sec:paragdims}

In Section \ref{sec:overview}, we briefly overview the smart grid.
In this section, we present four scenarios and the communication and
networking technologies that may be applicable in these scenarios.

\subsection{Smart Home}
\begin{figure}[h]
\centering
\includegraphics[width=0.5\textwidth]{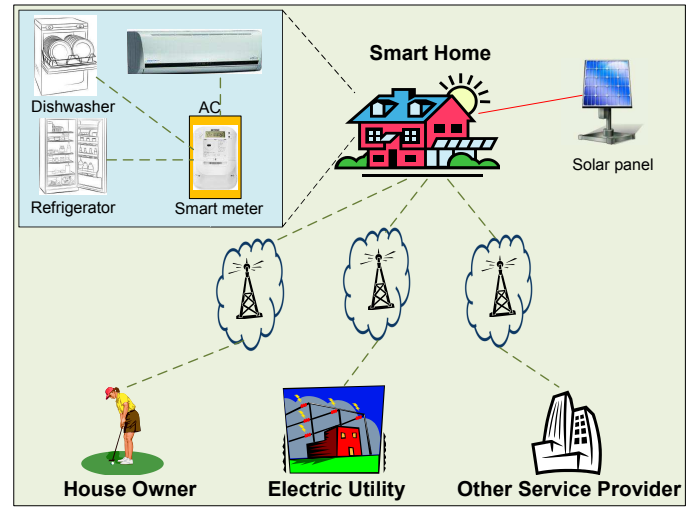}
\vspace{-0.1in} \caption{ 
An Example of Smart Home} 
\label{fig:smart_home}
\end{figure}

Automatic metering infrastructure (AMI) or  smart meters in the
smart grid enable two-way communications.
They have been widely regarded as the  most important mechanism  in
the smart grid  for obtaining information from end users' devices
and appliances, while also controlling the behaviors of the devices
\cite{Fang:Smart}.

A smart meter is usually an electrical meter that records
consumption in intervals of an hour or less and sends that
information at least daily back to the utility for monitoring and
billing purposes.
Since based on this infrastructure  all the information is available
in read time and on demand, both electric utility and users are able
to adjust the supply and demand profile to improve the system
operations and their own benefits.

In the future smart homes, many applications become possible based
on this infrastructure.

First,    end users are able to estimate bills and  thus manage
their energy consumptions to reduce bills.
For example, as shown in Fig.\ref{fig:smart_home}, the smart meter
can collect information and power consumption from the dishwasher,
the refrigerator, and the air conditioner.
Based on the realtime price information from electric utility, the
smart meter can  disconnect-reconnect remotely and control these
user appliances to reduce electricity cost.
In order to realize the wireless communications between the smart
meter and the user appliances,    ZigBee Smart Energy Profile (SEP)
has   been selected by a large number of utilities as the
communications platform \cite{Fang:Smart}.
This wireless communication standard provides a standardized
platform for exchanging data between smart metering devices and
appliances located on customer premises.
The features supported by the SEP include demand response, advanced
metering support, realtime pricing, text messaging, and load control
\cite{Yi:Developing}.

Second, the smart meter should also provide a communication and
control mechanism between smart home and remote house owner.
Let us consider a simple example pointed out by Xi \kETAL
\cite{Fang:Smart}.
In summer, the house owners hope that when they get home, the
temperature at home is around $60-80^\circ F$.
Thus, the smart meter connected to the air conditioner can
periodically inquire the position of the house owner by attempting
to send the inquiry information to the owner's smart phone which can
obtain the owner's position  via GPS.
If the smart meter finds the owner is coming back home, it will
decide to turn on the air conditioner in advance so that  when the
owner gets home, the temperature at home is around $60-80^\circ F$.
In this application, the smart meter is able to not only remotely
control the air conditioner via, for example, a Zigbee platform, but
also exchange information between house owner and itself.
For the latter information, the smarter meter may need to contact to
the service provider of the owner's cell phone via a cellular system
as shown in Fig.\ref{fig:smart_home}.

Third, in the future smart home, smart meter should not only be used
for an interface between the future smart home and electric utility.
Like ``Apple's Application Store''\cite{App:App}, many management
applications and services
 are available online.
Users can choose their expected services and download to the local
system (e.g. the smart meter) \cite{Fang:Smart}.
For example, a user who needs a management program supporting the
example about smart control of air conditioner mentioned above, can
buy this program from the Smart Grid Store online and download it.
This Smart Grid Store provides an integrated platform, which can
drive the third party to develop new management programs and
meantime help users easily customize their management services.

In addition to Zigbee and cellular systems, WIFI could also be used
in a smart home.
The most important advantage of WIFI is that it has been widely used
and has mature standards.
The smart meter can transmit its information to a WIFI access point,
and the information then would be routed to the electric utility.

\subsection{Microgrid}

\begin{figure}[h]
\centering
\includegraphics[width=0.3\textwidth]{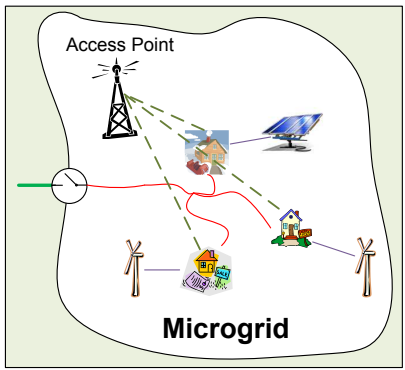}
\vspace{-0.1in} \caption{ 
An Example of Microgrid} 
\label{fig:microgrid}
\end{figure}

Microgrid is seen as one of the cornerstones of the future smart
grids \cite{Europe:SmartGrids, Fang:Online}.
The organic evolution of the smart grid is expected to come through
the plug-and-play integration of microgrids.
An important operation of microgrid is that   distributed  energy
generations and loads are grouped together, and thus the microgrid
can disconnect from the macrogrid and function autonomously.
This intentional islanding of generations and loads has the
potential to provide a higher local reliability than that provided
by the power system as a whole.

The information exchange among the users is needed for operating a
microgrid.
For example, first, they need to decide when they will operate in an
islanding mode, and how to optimize the power usage and resource
allocation if the microgrid is working in islanding mode.
Second, when the distributed generators (e.g. solar panels) of some
users generate more power than these users need, they may sell the
electricity to other users in this microgrid.
This transaction needs information exchange among these users.

A wireless mesh network is an applicable network architecture to
realize the information exchange among the users in a microgrid.
Note that a wireless mesh network   is a communication  network made
up of radio nodes organized in a mesh topology.
First, a wireless mesh network is a self-organized and
self-configured.
Considering that the smart grid allows a large number of
plug-and-play devices, this feature is very important.
Second, since usually multiple paths exist between any two nodes in
a mesh network, this provides a high communication reliability.
Third, mature research and industry standards have been carried out
for a wireless mesh network.
Wireless mesh networks can be implemented with various wireless
technology including IEEE 802.11, 802.15, 802.16, cellular
technologies or combinations of more than one type.


\subsection{Electric Vehicle System}
\begin{figure}[h]
\centering
\includegraphics[width=0.3\textwidth]{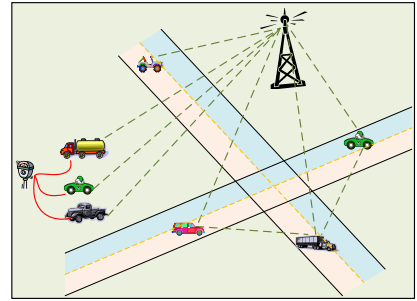}
\vspace{-0.1in} \caption{ 
An Example of Electric Vehicle System: Vehicles can exchange
information among them via a self-organized mobile adhoc network or
a base station.}
\label{fig:V2G}
\end{figure}

An electric vehicle is a vehicle that   uses one or more electric
motors for propulsion.
As   fossil fuels diminish and generally get more expensive,
electric vehicles or plug-in hybrid vehicles will rise in
popularity.
The wide use and deployment of electric vehicles will have a
significant impact on future grid.

The first impact is  that charging electric vehicles   will lead to
a significant new load on the existing distribution grids.
Recent study has shown that high penetration levels of uncoordinated
plug-in hybrid electric vehicles charging will significantly reduce
power system performance and efficiency, and even lead to
overloading \cite{Sortomme:Coordinated}.
%
%
The second impact is that electric vehicles provide a new way to
store and supply electric power.
This concept is called as Vehicle-to-Grid (V2G) in the vision of the
smart grid \cite{Fang:Smart}.
It allows V2G vehicles to provide power to help balance loads by
``peak shaving'' (sending power back to the grid when demand is
high) and ``valley filling'' (charging   when demand is low).

In order to integrate electric vehicle systems into power grid,
wireless information technologies will play an important role.

First, in order to mitigate the negative impact of electric vehicle
charging, the charging behaviors of   electric vehicles should be
regulated.
For example, it is recommended to do coordinated charging.
More specifically, a group of electric vehicles can optimize their
charging operations and  schedules based on their demands, to
guarantee that they will not charge at the same time or when the
power demand in the grid is already high.
Realizing the coordinated charging requires vehicles,  electric
utility, and even charging station to exchange a large amount of
information, such as location information, battery information,
charging service availability information, and grid power supply
information.
Based on this information, optimizing charging operations and
schedules is implementable.
Cellular communication systems   and mobile adhoc networks (MANETs)
are two   wireless communication architectures  which are applicable
in this scenario.
Cellular communication system    has already been used for vehicle
road assistance.
For example,  ``Google Map'' on smart phones provides real-time
traffic, which helps the users to find the fastest route.
Using existing cellular systems and smart phone platform to realize
this charging assistant programming has the following benefit.
%
%
Developing a charging assistant program on smart phone and using the
mature cellular system behind it as the information exchange medium,
instead of developing a new wireless system,  is a cost-effective
solution.
Every user can benefit from this technology as long as they install
this program on their smart phones.
In addition, MANET  is also a promising platform in this scenario.
MANET is a self-configuring infrastructureless network of mobile
devices connected by wireless links.
Such networks can further be connected to the larger Internet.
Based on MANET, vehicles can exchange the information efficiently
and thus make effective decisions.

Second, real-time information exchange is also needed for realizing
the concept V2G.
Suppose    an electric vehicle with excess power in a parking lot,
which has been connected to the   grid.
When the electric utility in the smart grid needs electric vehicles
to send the power back  to the grid in the peak time,  they must
  contact with the owner of this vehicle   to get a permission to use the battery of this
vehicle.
The most effective way to realize this information exchange is using
the existing cellular communication system.
After the utility gets the permission, it needs to remotely control
the vehicle.
This can be done via the  wireless network, where the vehicle is
located in.

\subsection{Monitoring System}

\begin{figure}[h]
\centering
\includegraphics[width=0.3\textwidth]{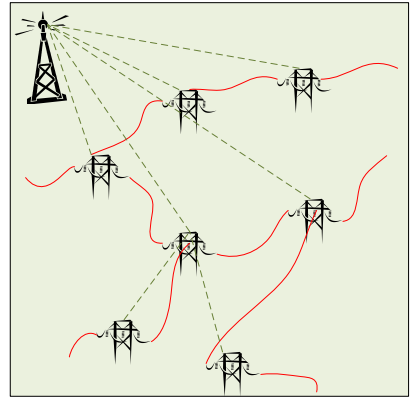}
 \caption{ 
An Example of Monitoring System} 
\label{fig:monitoring}
\end{figure}

Self-monitoring and self-healing are important features  in the
vision of the smart grid.
A large number of sensors are expected to be deployed in order to
detect failure events in power grids,  such as conductor failures,
tower collapses, hot spots, and extreme mechanical conditions.
Fig. \ref{fig:monitoring} shows an example of such monitoring
system, which uses wireless sensor networks to provide remote system
monitoring and diagnosis.
A base station  reports the grid status to the grid operator
periodically via a backbone communication network.
This backbone communication network is often built on a fiber optic
network.
The sensors collect the   status and transmit this information to
the base station via one-hop or multi-hop wireless networks.
%
Note that a wireless sensor network is probably the most
cost-effective way to realize this monitoring system for the
following reasons.
First, wireless sensors have fairly low installment and deployment
costs.
Second, if the density of the sensors is high enough, the sensor
networks would provide fairly high reliability and survivability.
This is because each area may be covered by multiple sensors.
Even if some sensors fail, it is highly likely that the important
areas are still being monitored.
This effectively reduces the maintenance cost.
%


\section{Challenges  for the Research on Wireless Technologies in Smart
Grid} \label{sec:challenge} In Section \ref{sec:paragdims}, we
presented the wireless communication and networking paradigms for
four important scenarios in the smart grid.
In this section, we   describe some issues and  challenges existing
in the research on the wireless  technologies for the smart grid.

\subsection{QoS and Entropy of Information Data}

The diversity of the information   transmitted within the smart grid
makes it necessary to differentiate the quality of service (QoS) of
the data.
Roughly speaking, we can categorize the data into two types:
critical data (e.g. the  critical grid status information collected
by the monitoring system) and non-critical data (e.g. user energy
consumption and billing information exchanged between the smart
meter and electric utility).
We must pay a particular attention on the QoS of the critical grid
status information.
In other words, we should guarantee that this type of data is sent
to the controller in a timely manner.
Otherwise, losing real-time grid monitoring may result in outage or
disastrous results, such as cascading blackouts.
Therefore, ensuring that the critical monitoring data is delivered
on time is a prerequisite to improve the grid reliability and
realize self-healing once failure events take place.
However, as the wireless environment is usually unreliable,
realizing this objective   is not easy task.
%

For the wireless communications in the smart grid, we also need  to
consider the entropy of the transmitted data.
Since a large number of sensors and smart meters are used in the
smart grid, a large amount of data (e.g. sensing data) will be
generated.
However, this data may have a large amount of redundancy.
For example, the smart meter readings must be similar when no
activity takes place at home.
The monitoring data generated by the sensors in the vicinity may
also have information redundancy.
It is well-known that the wireless spectrum is a  scarce resource.
Transmitting  a large amount of redundant data  significantly
reduces the  resource usage.
Therefore, improving the entropy of the transmitted data in the
smart grid would be beneficial.

\subsection{Wireless Communication Network Management and  Control}

A basic question about the wireless communication network used in
the smart grid   is: \emph{Should this network be   organized in a
distributed manner or a centralized manner}?
There is no straightforward answer to this question.
{\textbf{On one hand}}, the traditional electric utility will still
play a dominant role in the foreseeable future.
A  large amount of information should still be controlled and
managed by one or multiple large centralized electric utilities.
Note that although distributed wireless communication networks (such
as ad hoc networks) have been studied for a long time, it is
interesting that the centralized control structure is still  much
more popular, especially for the commercial systems where  central
utilities are involved.
{\textbf{On the other hand}}, many distributed entities are
introduced into the smart grid, which makes distributed
communication network be a competitive choice.
For example, a microgrid can be organized in a distributed manner.
It may prefer a distributed and self-organized communication network
structure, because a large number of plug-and-play devices may be
used and a microgrid is expected to have the capacity of functioning
autonomously.
%
%
%
%
In brief, how to effectively organize the wireless network in the
smart grid is still an open question and is worth further
investigation.

Another question is how to   optimize the complicated heterogenous
wireless communication system underlying the smart grid, where
multiple wireless communication technologies are used
simultaneously.
Note that due to the industry standards and utility interests these
technologies may overlap in both time and space.
For example, a smart meter may exchange data with the user
appliances via Zigbee (due to the user appliance industry standard),
transmit information  to the electric utility via WiMax (because it
is   cost-effective for the electric utility to set up or lease one
base station to cover a large number of users), and keep  in touch
with the house owner's phone via a cellular communication system
(suppose that the house owner always carries a 3G smart phone).
The question is how to jointly optimize these different wireless
networks to reduce cost and improve the wireless resource
utilization.
%

\subsection{Security and Privacy}

Cyber security is regarded as one of the biggest challenges in the
smart grid \cite{NIST:NIST}.
The malicious attacks on the wireless communication networks
underlying the smart grid can be categorized into   three major
types based on their goals \cite{Lu:Review, Fang:Smart}: network
availability, information privacy, and data integrity.

\emph{Network Availability}: Malicious attacks targeting network
availability   attempt to delay or block  information transmission
in order to make system resources unavailable to nodes that need to
exchange information in the smart grid.
As a result, the real-time monitoring of critical power
infrastructures may be lost, which may further lead to a possible
global power system disasters.
For example, in power grids cascading failure is common  when one of
the elements fails  and shifts its load to nearby elements in the
system.
Those nearby elements are then pushed beyond their capacity.
As a result, they become overloaded and shift their load onto other
elements.
This failure process cascades through the elements of the system
 and continues until substantially all of the elements in the system are compromised.
In concept, smart grid is expected to handle this case by widely
deploying monitoring devices to monitor the grid realtime status.
When wireless communication network availability is compromised, the
monitoring information cannot be transmitted effectively  to the
controller.
As a result, one element failure may lead to a disastrous cascading
failure.
Therefore, as pointed out by National Institute of Standards and
Technology (NIST) \cite{NIST:NIST}, the design of information
transmission networks that are robust to attacks targeting network
availability is the top priority.

\emph{Information privacy}: The major benefit provided by the smart
grid, i.e. the ability to get richer data to and from customer
meters and other electric devices, is also its Achilles' heel from a
privacy viewpoint \cite{NIST:NIST}.
The energy use information stored at the meter acts as an
information-rich side channel.
This opens up a door for a malicious attacker, who is interested in
the personal information such as individual's habits, behaviors,
activities, preferences, and even beliefs.
Once the wireless transmission security is broken by the attacker,
the attacker can retrieve this privacy information  by analyzing
energy use information.

\emph{Data integrity}: Data integrity attack attempts to
deliberately modify or corrupt information transmitted within the
smart grid.
This may lead to an extremely damaging result in the smart grid.
For example, Liu \kETAL  \cite{Liu:False} showed that an attacker
can manipulate the state estimate without triggering bad-data alarms
in the control center.
This attack is called the false-data injection attack or the stealth
attack.
Consider the cascading failure example above.
An attacker can compromise one grid element and prevent the control
center from detecting this by using stealth attack.
This may result in a disastrous cascading failure.

In summary, the advanced communication infrastructure introduced in
the smart grid is a double-edged sword.
On one hand, it provides a way for the power grid to realize
complicated  operations and functions.
On the other hand,  expanded communication paths   can easily result
in an increase in vulnerability to cyber attacks and system
failures.
Security is a never-ending game of wits, pitting attackers versus
asset owners.
How to guarantee cyber security for the smart grid will be a
long-term research topic.

\subsection{Interoperability and Compatibility}
Thus far, the smart grid is not a thing but rather a vision.
It is regarded as a loose integration of complementary components,
functions, subsystems, and services under the pervasive control of
highly intelligent management-and-control systems.
Therefore, many different wireless communication protocols and
technologies probably will be used in the smart grid.
%
%
As a result, realizing interoperability among them is not an easy
task.
Although NIST \cite{NIST:NIST} has proposed a draft of framework and
roadmap for smart grid interoperability standards, many problems are
still left to address.

First, we need to consider how to  take advantage of the legacy
infrastructure, in order to reduce the deployment and installment
cost of the wireless communication system underlying the new smart
grid.
The cellular communication system and the WIFI network are two
widely used wireless infrastructure.
It would be beneficial to investigate how to effectively integrate
these two systems into the smart grid vision.

Second, we should  consider not only how to take advantage of the
existing wireless communication systems, but also the system forward
compatibility.
Since the current concept of the smart grid is just a vision, many
new features will be integrated into the new grid in the future.
We must consider that the current design for the wireless
communication systems underlying the smart grid  can also be used or
can be easily upgraded to support new features.
%

Third, from a technical point of view,  the classic layer model
(e.g. the famous Open Systems Interconnection model) could provide a
promising conceptual solution to realize interoperability among
different wireless communication technologies.
However, it is well-known that this layer model suffers in some
modern applications.
For instance, the performance of the pure TCP may be very bad in
wireless networks, because  it cannot differentiate  packet loss due
to wireless fading from that due to a real congestion in the
network.
Therefore, in practice some functions or services  are not tied to a
given layer, but   can affect more than one layer, in order to
improve the quality of services.
This  concept  often requires {cross-layer design and optimization}.
%
%
%
However,   interoperability among different communication
technologies,   a precursor to cross-layer approaches, is difficult
\cite{Fang:Smart}.
%

%
%

\section{Conclusions} \label{sec:conc}

There is no doubt that the smart grid will lead to better power
supply services and a more environmentally sound future.
However,  we   still have a long way to go before this vision comes
true.
Communication system is a nervous system of this new grid and
requires a large amount of research effort.
In the article, we have visioned  the wireless communication and
networking paradigms for four typical scenarios in the smart grid
and point out the research challenges for wireless communication and
networking technologies in the smart grid.
We believe that more and more communication paradigms would emerge
as the research on the smart grid is extended.
Eventually, the information transmission network and energy delivery
system will be organically integrated, which will revolutionize our
daily life.




\begin{IEEEbiography}[{\includegraphics[width=1in,height=1.25in,clip,keepaspectratio]{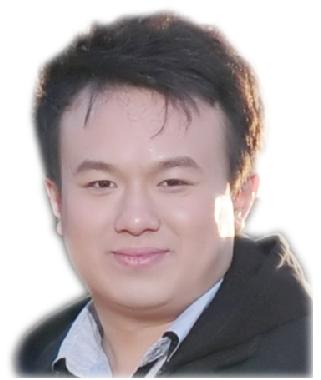}}]{Xi Fang}
(Student Member 2009) received the B.S. and M.S. degrees from
Beijing University of Posts and Telecommunications, China, in 2005
and 2008 respectively. Currently he is a Ph.D student in the School
of Computing, Informatics, and Decision Systems Engineering at
Arizona State University. He   has received a Best Paper Award at
{\sf IEEE MASS'2011 (IEEE International Conference on Mobile Ad-hoc
and Sensor Systems)}, a Best Paper Award at {\sf IEEE ICC'2011(IEEE
International Conference on Communications)}, and a Best Paper Award
Runner-up at {\sf IEEE ICNP'2010 (IEEE International Conference on
Network Protocols)}.  His research interests include algorithm
design and optimization in wireless networks, optical networks,
cloud computing, and power grids.
\end{IEEEbiography}

\begin{IEEEbiography}[{\includegraphics[width=1in,height=1.25in,clip,keepaspectratio]{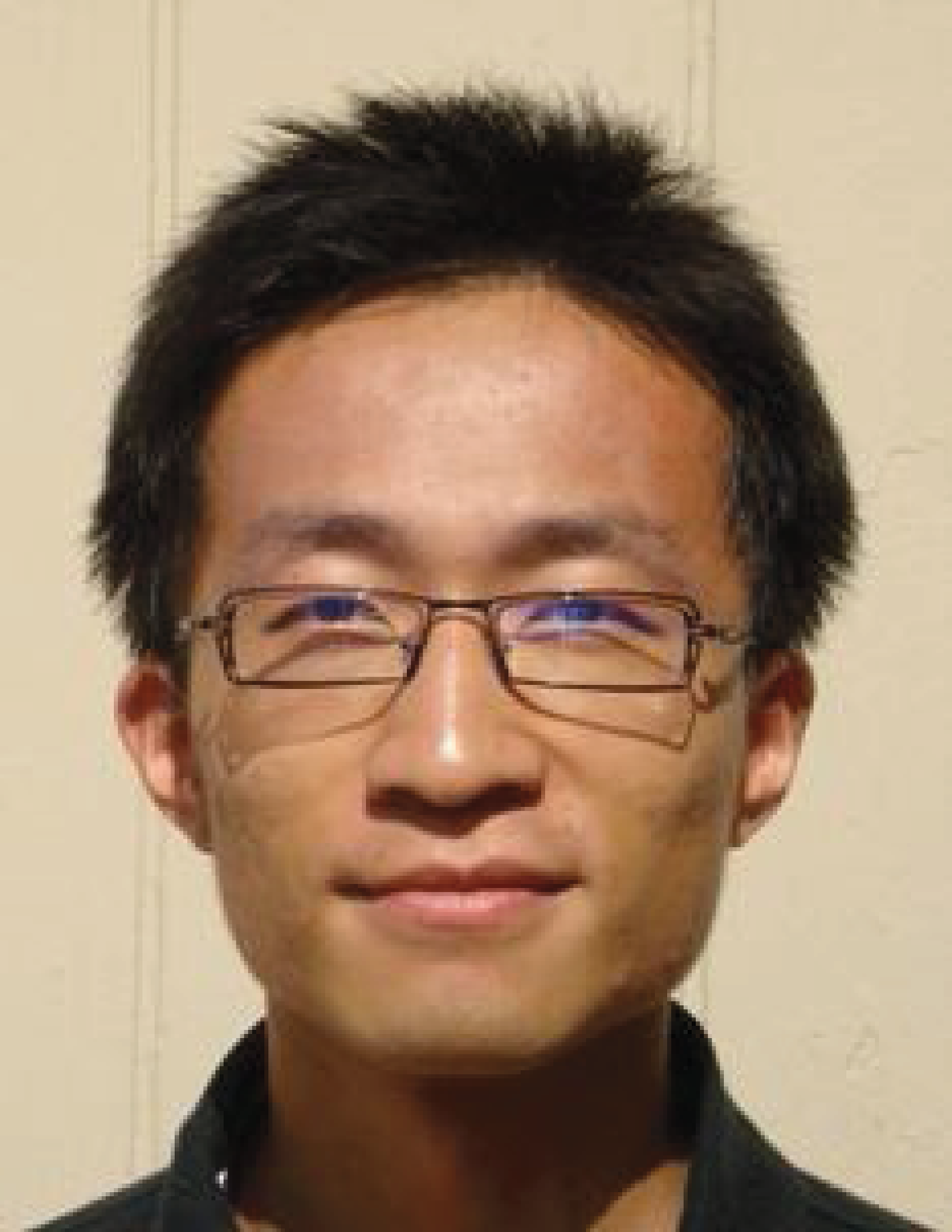}}]{Dejun Yang}
(Student Member 2008)  received his B.S. from Peking University,
Beijing, China, in 2007. Currently he is a PhD student in the School
of Computing, Informatics, and Decision Systems Engineering (CIDSE)
at Arizona State University. His research interests include economic
and optimization approaches to wireless networks and social
networks. He has published over 20 papers in venues such as {\sf
IEEE Journal on Selected Areas in Communications}, {\sf IEEE
Transactions on Computers}, {\sf IEEE Transactions on Mobile
Computing}, {\sf ACM MOBIHOC}, {\sf IEEE INFOCOM}, {\sf IEEE ICNP},
{\sf IEEE ICC}, and {\sf IEEE Globecom}.
He has received a Best Paper Award at {\sf IEEE ICC'2011}, a Best
Paper Award at {\sf IEEE MASS'2011} and a Best Paper Award Runner-up
at {\sf IEEE ICNP'2010}.
\end{IEEEbiography}

\begin{biography}[{\includegraphics[width=1in,height=1.25in,clip,keepaspectratio]
{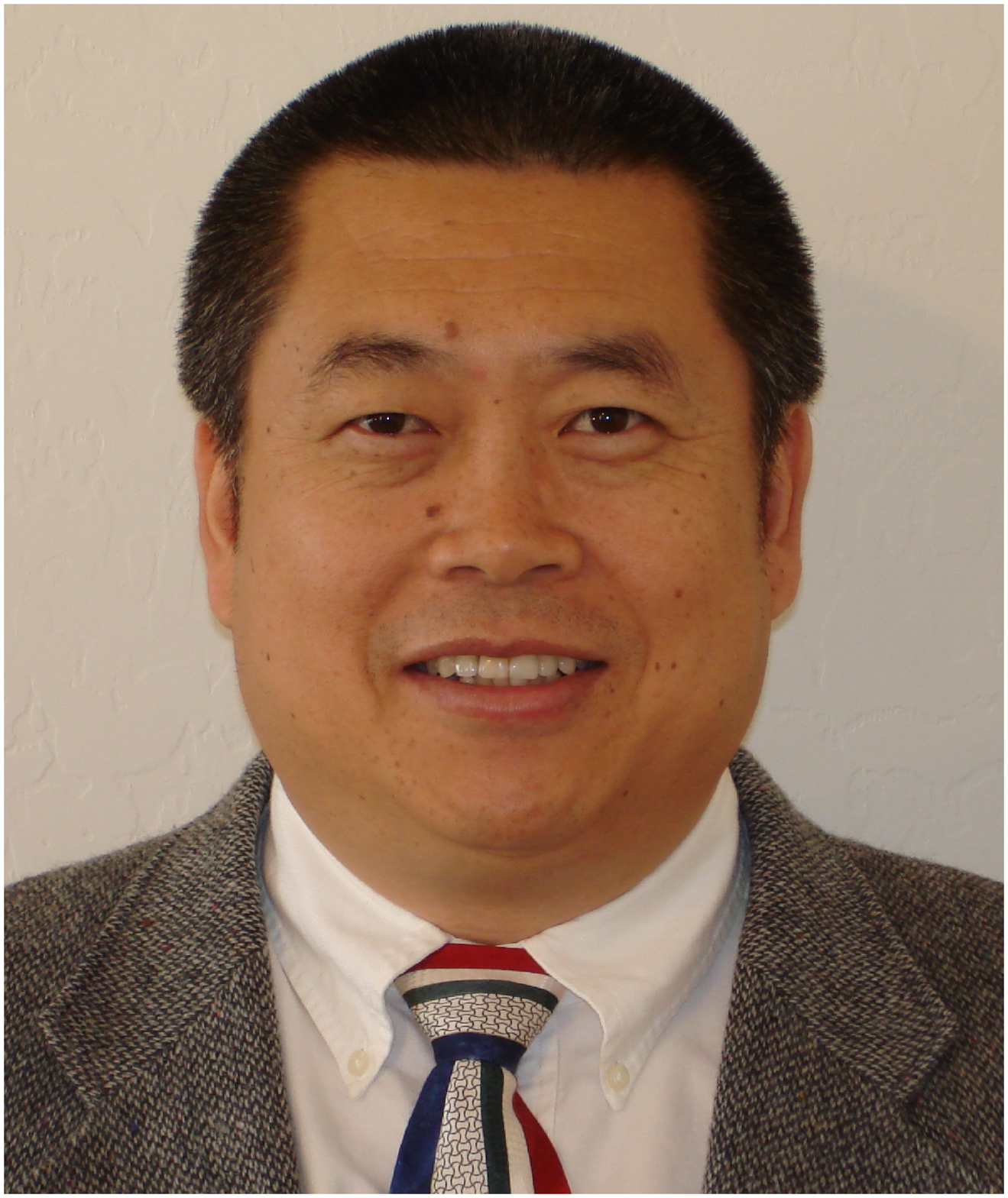}}]{Guoliang Xue}(Member 1996, Senior Member 1999, Fellow,
2011) is a Professor of Computer Science and Engineering at Arizona
State University.
He received the BS degree (1981) in mathematics and the MS degree
(1984) in operations research from Qufu Normal University, Qufu,
China, and the PhD degree (1991) in computer science from the
University of Minnesota, Minneapolis, USA.
His research interests include survivability, security, and resource
allocation issues in networks ranging from optical networks to
wireless mesh and sensor networks and social networks.
He has published over 200 papers (including 97 journal papers) in
these areas.
He received a Best Paper Award at IEEE Globecom'2007 (IEEE Global
Communications Conference), a Best Paper  Award Runner-up at IEEE
ICNP'2010 (IEEE International Conference on Network Protocols), and
a Best Paper Award at IEEE ICC'2011 (IEEE International Conference
on Communications).
He received the NSF Research Initiation Award in 1994, and has been
continuously supported by federal agencies including NSF and ARO.
He is an Associate Editor of {\sf IEEE/ACM Transactions on
Networking} and {\sf IEEE Network} magazine, as well as an Editorial
Advisory Board Member of {\sf IEEE Transactions on Wireless
Communications}.
He served as an Associate Editor of {\sf IEEE Transactions on
Wireless Communications} and {\sf Computer Networks} journal.
His recent conference services include TPC co-chair of {\sf IEEE
INFOCOM'2010}, Symposium co-Chair of {\sf IEEE ICC'2009}, General
co-Chair of {\sf IEEE HPSR'2008}, and Panels Chair of {\sf ACM
MOBIHOC'2008}.
He was a Keynote Speaker at {\sf LCN'2011 (the 36th IEEE Conference
on Local Computer Networks)} and a Plenary Speaker at {\sf
ICCCN'2011 (International Conference on Computer Communication
Networks)}.
He is a {\sf Distinguished Lecturer} of the IEEE Communications
Society.
He is a Fellow of the IEEE.
\end{biography}


\end{document}